\documentclass[12pt]{article}
\usepackage{epsfig}
\begin{document}

\vspace*{5.00 cm}

\centerline{\bf\Large{The case for a cosmological constant}} 

\vspace*{2.00 cm}

\centerline{\bf Matts Roos and S. M. Harun-or-Rashid} 

\vspace{2.00 mm}

\centerline{Department of Physics, Division of High Energy Physics, 
University of Helsinki, Finland}

\newpage

\centerline{\bf\Large{The case for a cosmological constant}}

\vspace*{1.00 cm}

\centerline{Matts Roos and S. M. Harun-or-Rashid} 
\centerline{Department of Physics, Division of High Energy Physics,} 
\centerline{University of Helsinki, Finland.} 

\vspace*{1.00 cm}
Taking the Hubble constant to be in the range 60 - 75 km/s Mpc 
we show that three independent conditions strongly rule out the
standard model of flat space with vanishing cosmological constant. \\

\section{Introduction}
The dynamical parameters describing the cosmic expansion are still very
inaccurately known, permitting a wide range of cosmological models to appear
plausible. Judging from all the information available at present, the Universe
is either open, with the total density parameter $\Omega_0<1,$ or flat by
virtual of the presence of a cosmological constant. Let us use the  notation 
\begin{eqnarray}
\Omega_0 = \Omega_m + \Omega_{\lambda}=1\,
\label{f1}\end{eqnarray}
where the density parameter of the vacuum energy is $\Omega_{\lambda} =
{\lambda}/3H_0^2$, and $\Omega_m$ is the density parameter of baryonic + dark
matter.

We study here whether the ``standard model'' of flat space with $\Omega_\lambda
=0$ is still viable in the light of recent determinations of several
cosmological parameters.

\section{Cosmological conditions}
The Hubble constant $H_0$ used to be uncertain by a factor of two, but it now
is converging into the (one standard deviation) range
\begin{eqnarray}
H_0 = 67 \pm 6 \rm{\ km \ s^{-1} Mpc^{-1}}. 
\label{f2}\end{eqnarray}
\noindent This value is the result by Nevalainen $\&$ Roos (1997) of combining
$H_0$ determinations from four HST-observed galaxies, and applying the
correction due to the metallicity dependence of the Cepheids, as determined by
Beaulieu \& al. (1997) and Sasselov \& al. (1997), to Cepheids in M96 (Tanvir
\& al. 1995) and M100 (Freedman \& al. 1994) and to the supernov\ae NGC 5253
(Sandage \& al. 1994) and IC 4182 (Saha \& al. 1994, Sandage \& al. 1994).
If one takes into account the correction to the Cepheid period-luminosity
relation measured by Hipparcos (Feast \& Catchpole 1997) the above $H_0$
value may still go down by 10\%, but we feel that it is still not quite settled
whether this is justified at the distances in question.

An age limit of the Universe can be taken from the age of the oldest
globular clusters. Making use of the new distance measurements by Hipparcos, 
Chaboyer \& al. (1997) estimate their mean age to be 
\begin{eqnarray}
t_{globulars} = 11.5\pm 1.3\ \rm{Gyr}\,  
\label{fn}\end{eqnarray}
To which the unknown age of the Universe at the time of their formation must be
added. An even more stringent limit is posed by the discovery
(Dunlop \& al. 1996, Kashlinsky \& Jimenez 1996) of a weak and extremely red
radio galaxy 53W091 at $z=1.55$ whose spectral data indicate that its stellar
population is at least 3.0 Gyr old. Although Bruzual \& Magris (1997) have
claimed that this age determination is an artifact, we shall explore the
consequences of such an old object.

In the Friedman-Lema\^{\i}tre model the age $t(z)$ of the Universe
at redshift $z$ can be expressed in the form
\begin{eqnarray}
t(z)&=&{{1}\over
{H_0}}\int_0^{1/(1+z)}\hbox{d}x\big{[}(1-\Omega_m-\Omega_{\lambda})\nonumber \\
&+&\Omega_mx^{2-3(1+\alpha)} +\Omega_{\lambda}x^2 \big{]}^{-1/2}.
\label{f3}\end{eqnarray}
Here $\alpha$ is the ratio of pressure to energy density, thus it defines the
equation of state. For ordinary or dark non-relativistic pressureless matter 
$\alpha=0$. For a flat, static Universe with a cosmological constant 
$\alpha=-1$.

Using the Dunlop \& al. (1996) age of the radio galaxy 53W091, 
\begin{eqnarray}
t(z)=t(1.55) = 3.5\ \rm{Gyr}\,
\label{f4}\end{eqnarray}
Eq.(\ref{f3}) constrains the space of $H_0,\ \Omega_m,\ \Omega_{\lambda}$ and
$\alpha$. In figure 1 we plot the solution to Eq.(\ref{f3}) for $\alpha =0$ and
the limit (4) for several values of $H_0$. It is obvious that the standard
corner  $(\Omega_{\lambda}=0, \Omega_m=1)$ is then strongly disfavoured by the
$H_0$ value (2).

The only way to reduce the age below 3.0 \rm{Gyr} is to reduce the product
$\Omega_mH_0$, but this then reduces the small scale power in the primordial
density field beyond allowed limits. One degree of freedom which can be used to
improve this situation has been pointed out by Steinhardt (1996). Although
$\alpha$ is traditionally taken to be zero, in a universe with interacting
fields and topological defects $\alpha$ can vary between -1 and 0. This implies
the presence of strings causing more small scale power than in simple
inflation. In Figure 2 we plot the situation with $\alpha=-0.1$ for different
$H_0$-values. This moves the allowed region towards the standard corner.
However, the price paid for this improvement may seem too high, also in view of
the next conditions below.

Let us alternatively look at the consequences of taking $t_0$ to be given
by the age of the globular clusters. The Chaboyer \& al. (1997) determination
carries a $1\sigma$ error of 1.3 Gyr. Let us assume that the age of the
Universe at the time of their formation is short enough to be included in this
error, then we may take
\begin{eqnarray}
t_0\approx 11.5 + 1.3 \ \rm{Gyr}=12.8\ \rm{Gyr}.
\label{fa}\end{eqnarray}

\noindent Requiring Eq.(\ref{f3}) to yield this value when $H_0$ has the value
in Eq.(\ref{f2}), one finds that
\begin{eqnarray}
\Omega_m = 0.40\ ^{+0.15}_{-0.10}
\label{fb}\end{eqnarray}
in a flat Universe. Thus a considerable $\Omega_{\lambda}$-component is
required. In an open Universe with $\Omega_{\lambda}=0$, $\Omega_m$
would be very small, 0.3 at most.

The second condition ruling out the standard  model is the observational value
of $\Omega_m$ as determined by X-ray studies of gas in clusters, e.g. the Coma
cluster (White \& al. 1993) and the rich cluster A85 (David \& al. 1995,
Nevalainen \& al.  1997). Assuming that the gravitating matter seen in these
galaxies out to a radius of about 3 Mpc extrapolate well to the average matter
density of the Universe, one can obtain a value for the quantity 
$(\Omega_B/\Omega_m)h^{3/2}$. To evaluate $\Omega_m$ one needs to know
the baryonic density parameter $\Omega_B$ which is highly controversial, due
to the conflicting deuterium observations (Rugers \& Hogan 1996, Tytler \& al.
1996, 1997, Webb \& al. 1997). Pushing the gas density profile parameters to
their $90\%$ confidence limit, and the Hubble parameter at its lower
(two-sided) $3\sigma$ limit, $H_0 \ge 50 \rm{\ km \ s^{-1}Mpc^{-1}}$, and taking
the D/H ratio to be maximal (Webb \& al. 1997), on obtains
\begin{eqnarray}
\Omega_m \le {0.22} \ . 
\label{f5}\end{eqnarray}

Large-scale structures (galaxy correlation functions, the abundance of rich
clusters, cluster-cluster correlations etc.) do not constrain $\Omega_m$ very
tightly, but in the $H_0$ range (2) they prefer values larger than
Eq.(\ref{f5}), $\Omega_m \ge 0.3$ or so. Thus there is some conflict but in no
way driving the solution into the standard corner. Perhaps the solution to this 
conflict is anti-biasing, or perhaps the X-ray studies of gas in clusters
reflect local conditions which are different from cosmological values.

Plotting the limit (\ref{f5}) in Figures 1 and 2 it is obvious that it strongly
rules out the standard corner, but it can be made to agree with a flat
$\Omega_0=1$ universe if the vacuum energy density contribution is large.

The third condition ruling out the standard model is due to strong and weak 
gravitational lensing. From observations of the clusters Cl $0024+1654$ and
A370, Mellier \& al.(1997) conclude that 
\begin{eqnarray}
\Omega_{\lambda} \ge 0.6.
\label{f7}\end{eqnarray}

\noindent Note that previous lensing studies only gave upper limits, {\it e.g.}
Kochanek (1996) found 
\begin{eqnarray}
\Omega_{\lambda}\le 0.66
\label{fc}\end{eqnarray}
for a flat Universe.

{\noindent} As can be seen in the Figures, these limits are in agreement with
the  limit (\ref{f7}) and the $H_0$ range (\ref{f2}) when taking the age of the
Universe from the  53W091. It also agrees roughly with the upper limit to
$\Omega_{\lambda}$ from the observations of Perlmutter \& al (1997) of the
light-curves of seven high redshift supernov\ae. 
When $t_0$ is taken from the globular clusters, these conclusions are
somewhat softened, but qualitatively the same.

We have not made any use of CMB data, because the height of the Doppler peak
has not been measured well enough yet to yield information of precision
comparable to what was used here. Note that both large-scale structures and the
CMB Doppler peak depend on further adjustable parameters which we have not
referred to, such as bias $b$ and the spectral index $n$.

\section{Conclusion}Thus our conclusion is that the standard model with
$\Omega_{\lambda}=0$, $\Omega_m =1$ is ruled out, some low-density open models
are possible, but the preferred range is around $\Omega_{\lambda}=0.6-0.8,$
$\Omega_m = 0.2 - 0.4$, $H_0 = 60-75 \rm{\ km \ s^{-1} Mpc^{-1}}$  and the
flat geometry of Eq.(\ref{f1}) is possible. 
\begin{figure}
\centering
\epsfig{file=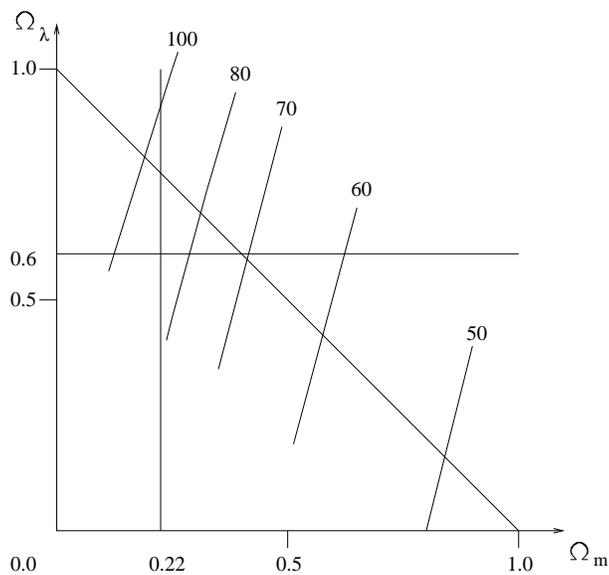,angle=0,width=0.5\textwidth}
\caption{Regions allowed in the $(\Omega_0, \Omega_{\lambda})$-space. The set 
of solid lines are imposed by the observation of the radio galaxy 53W091
assuming $\alpha=0$. The numbers indicate the value of the Hubble constant in
units of km s$^{-1}$ Mpc$^{-1}$. The limits $\Omega_m <0.22$, 
$\Omega_{\lambda}>0.6$ are indicated. }
\label{Fig1}
\end{figure}

\begin{figure}
\centering
\epsfig{file=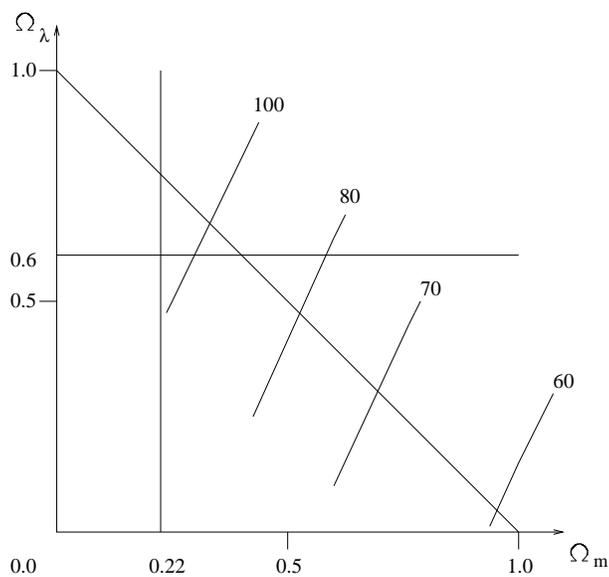,angle=0,width=0.5\textwidth}
\caption{As Fig. 1, but for $\alpha$ = -0.1}
\label{Fig2}
\end{figure}
\break 

\section*{References}
Beaulieu J. P. \& al., 1997, A\&A 318, L47\\
Bruzual G. \& Magris G., 1997, Preprint astro-ph/9707154
Chaboyer B. \& al., 1997, Preprint astro-ph/9706128, Subm. to APJ \\
David L. P. \& al., 1995, ApJ 445, 578-590 \\
Dunlop J. \& al., 1996, Nature 381, 581\\
Feast M. W. \& Catchpole R. M., 1997, MNRAS 286, L1\\
Freedman W. \& al., 1994, Nature 371, 757\\
Kashlinsky A. \& Jimenez R., Preprint astro-ph/9610269\\
Mellier Y. \& al., 1996, Preprint astro-ph/9609197\\
Nevalainen J. \& Roos M., 1997, Subm. to A\&A\\
Nevalainen J. \& al., Proc. 2nd Integral Workshop, ESA Sp-382, (1997) (in
press)\\ 
Perlmutter S. \& al., 1997, ApJ 483, 565\\
Rugers M. \& Hogan C. J., 1996, AJ 111, 2135\\
Saha \& al., 1994, ApJ 425, 14\\
Sandage A. \& al., 1994, ApJ 423, L13\\
Sasselov D. D. \& al., 1997, astro-ph/9612216 and A\&A\\
Steinhardt P. J. \& al., 1996, Nature 382, 768\\
Tanvir N. R. \& al., 1995, Nature 377, 27\\
Tytler D., Fan X.-M. \& Burles S., 1996, Nature 381, 207\\
Tytler D.,  Burles S. \& Kirkman D., 1997, Preprint astro-ph/9612121, Subm. to APJ\\
Webb J. K. \& al., 1997, Nature 388, 250\\
White S. D. M. \& al., 1993, Nature 366, 429-433\\ 
Winkler C. \& al., 1997, ESA Sp-382, (in press)\\

\end{document}